# Silver environment and covalent network rearrangement in GeS$_3$-Ag glasses

L. Rátkai[1], I. Kaban[2,3], T. Wágner[4], J. Kolář[4], S. Valková[4], Iva Voleská[4], B.Beuneu[5], P. Jóvári[1]

[1] Wigner Research Centre for Physics, Institute for Solid State Physics, H-1525 Budapest, POB 49, Hungary

[2] TU Dresden, Institute of Materials Science, D-01062 Dresden, Germany

[3] IFW Dresden, Institute for Complex Materials, P.O. Box 270116, D-01171 Dresden, Germany

[4] University of Pardubice, Faculty of Chemical Technology, Centre of Materials and Nanotechnology, Čs. legií 565 square, 53210 Pardubice, Czech Republic

[5] Laboratoire Léon Brillouin, CEA-Saclay, 91191 Gif sur Yvette Cedex, France

**Abstract**

The structure of Ag-doped GeS$_3$ glasses (0, 15, 20, 25 at.% Ag) was investigated by diffraction techniques and extended X-ray absorption fine structure measurements. Structural models were obtained by fitting the experimental datasets simultaneously by the reverse Monte Carlo simulation technique. It is observed that Ge has mostly S neighbours in GeS$_3$, but Ge-Ge bonds appear already at 15% Ag content. Sulphur has ~2 S/Ge neighbours over the whole concentration range, while the S-Ag coordination number increases with increasing Ag content. Ag-Ag pairs can already be found at 15% Ag. The Ag-S mean coordination number changes from 2.17±0.2 to 2.86±0.2 between 15% and 25% Ag content. Unlike As-S network in AsS$_2$-25Ag glass, the Ge-S network is not fragmented upon Ag-doping of GeS$_3$ glass.

## 1. Introduction

Memristive switching [1] has attracted high attention in view of development of advanced nonvolatile memories. There exist several concepts of resistance-change memory such as for example electric switching in chalcogenide glasses [2], resistive switching in transition-metal oxides [3,4], and resistance change in solid electrolytes [3-8]. The first concept utilizes amorphous-to-crystalline phase-change, while the second one uses valency change of transition-metal oxide upon applied electrical pulse. The third concept is based on nanoscale ionic transport and electrochemical formation or removal of nanoscale conductive pathways in solid-electrolyte matrix depending on polarity of electrodes. Devices utilizing this phenomenon are called electrochemical metallization cells (EMC) or conductive bridge memories (CBM).

Different candidate materials for application in the EMC cells are under investigation at present [3-13]. Among them are Ge-Se or Ge-S chalcogenide glasses as solid electrolyte containing Ag or Cu as active metal [5-7]. In comparios to the Ge-Se-Ag glasses, Ag-doped Ge-S glasses are able to withstand much higher temperatures, which is essential for providing long term resistance-switching functionality. Therefore, it appears to be a better candidate for EMC memory applications. To optimize their functionality it is important to understand the structure of Ge-S-Ag glasses. In this paper we present the results of a study on $GeS_3$-Ag glasses with 0, 15, 20, 25 at.% Ag. Atomic level structural models are generated by fitting diffraction and extended X-ray absorption fine structure (EXAFS) datasets with the reverse Monte Carlo (RMC) simulation technique [14-16]. This method offers a frame for combining experimental structural information with physical/chemical knowledge (e.g. density, preferred coordination numbers) available a priori. The validity of various structural models can also be tested. This approach has already been applied to several closely related systems such as $AsS_2$-Ag [17] and $As_2S_3$-Ag [18], $GeSe_3$-Ag [19] and Ge-Se-In glasses [20].

## 2. Experimental

### 2.1. Sample preparation

Four glasses of the composition $(GeS_3)_{100-x}Ag_x$ with x = 0, 15, 20, 25 at.% were prepared from 5N purity elements. The components of total mass of 10 g were inserted into the quartz ampoules, which were then evacuated to a pressure of $10^{-3}$ Pa, sealed, and placed in a rocking furnace. The glasses were synthetized at the well defined heating profile with a maximum temperature of 1000°C kept for 12 hours. The ampoules were quenched in iced water and then these ampoules were annealed for 3 hours at 50 K below the respective glass transition temperature.

The mass density of the glasses was determined with accuracy of 0.15% by dual weighing by the standard Archimedean method at room temperature. The measured values are given in Table 1.

### 2.2. Experiments

Neutron diffraction measurements were carried out at the 7C2 diffractometer of Laboratoire Léon Brillouin (Saclay, France). Powder samples were filled into vanadium sample holders with 0.1 mm wall thickness and 6 mm diameter. The wavelength of incident radiation was 0.72 Å. Raw data were corrected for background scattering and detector cell efficiency and normalized following standard procedures.

Ge and Ag K-edge EXAFS spectra were recorded at beamline X1 of Hasylab, Hamburg. Measurements were carried out in transmission mode. Powder samples were mixed with cellulose and pressed into tablets. The transmission of the tablets was around 1/e at the measured absorption edge. Intensities before and after the samples were recorded with ionisation chambers filled with Ar and Kr with pressures depending on the energy of the edge. The X-ray absorption cross sections $\mu(E)$ were converted to $\chi(k)$ by the program Viper [21].

X-ray diffraction data were taken at the BW5 high energy X-ray diffractometer [22] (Hasylab). The energy of incident radiation was 100.0 keV ($\lambda$ = 0.124 Å). Raw data were corrected for background scattering, detector deadtime, and Compton scattering [23].

**3. Reverse Monte Carlo simulation**

Large scale structural models (12000 atoms) were obtained by fitting simultaneously diffraction and EXAFS datasets. Simulations were carried out with the rmcppm code [16]. Number densities used in the models were calculated from the experimental mass densities (table 1). Minimum interatomic distances (cut offs) are listed in table 2. Backscattering amplitudes and phases needed to transform partial pair correlation functions to model $\chi(k)$ curves [24] were obtained by the feff programme [25]. Besides, minimum intertomic distances coordination constraints were also used: each Ge atom was forced to have four neighbours (either S or S/Ge, see below), each S atom had at most two Ge neighbours and no 'floating' Ag or S atoms (with 0 or 1 neighbours) were allowed.

**4. Results and discussion**

*4.1. $GeS_3$ glass*

Structural model of glassy $GeS_3$ was obtained by fitting simultaneously ND, XRD and Ge K-edge EXAFS datasets. As this composition is sulfur-rich, Ge-Ge bonding was forbidden while S-S bonding was allowed during the simulation. For this composition only, the coordination number of sulfur was constrained to be 2.

Experimental curves and fits are compared in figure 1 while partial pair correlation functions are shown in figure 2. The Ge-S distance is 2.23 Å which agrees well with Ge-S bond lengths (2.20-2.23 Å) found in $GeS_2$ [26] and in Ge-In-S, Ge-In-S-AgI and $Na_2S$-$GeS_2$ glasses [27-29]. The S-S distance is 2.06 Å, which is close to the value found in amorphous sulfur [30].

*4.2. $GeS_3$-Ag glasses*

The structure of ternary $GeS_3$-Ag alloys was investigated by fitting the four measurements (XRD, ND, and EXAFS at Ge and Ag absorption edges) simultaneously. The quality of the fits is demonstrated in figure 3, while coordination numbers and nearest neighbour distances are summarized in tables 3 and 4. Selected partial pair correlation functions are shown in figure 4.

Previous studies on $AsS_2$-Ag [17], $GeSe_3$-Ag [19] and $GeS_2$-$Ag_2S$ [31] glassy systems revealed that Ag prefers S/Se and tries to avoid the network forming cation (As/Ge). Our results show that from this point of view $GeS_3$-Ag alloys behave in a similar way. Ag has on the average 2.17±0.2 sulphur neighbours in $GeS_3$-15Ag, while it is coordinated by 2.86±0.2 sulfur atoms in $GeS_3$-25Ag. The Ag-S distance is 2.53-2.57 Å, which agrees with the Ag-S bond lengths found in $As_2S_3$-$Ag_2S$ [32], but it is slightly shorter than the values found in $GeS_2$-$Ag_2S$ (2.58 Å [31]) and in $GeS_2$-$In_2S_3$-AgI glasses (2.60 Å [33]). The deviations may be caused partly by experimental uncertainties, partly by the composition dependence of Ag-S interaction. The Ag-Ag distance is around 3.0 Å. A similar value was found in $As_2S_3$-$Ag_2S$ [32]. On the other hand, a neutron diffraction study on isotopically substituted $GeS_2$-$Ag_2S$ gave no clear conclusion: while a peak of $g_{AgAg}(r)$ was reported around 3 Å in the RMC study of Lee et al [34], direct transformation of diffraction datasets gave no evidence of nearest neighbour silver atoms [31]. It is to be mentioned that the Ag-Ag coordination number (0.73-0.93) does not change significantly in the composition range investigated.

With increasing Ag content, the initially sulfur-rich $GeS_3$-Ag system can turn to a sulfur-deficient state where Ge-Ge bonding may be required to satisfy the four valences of germanium. Test calculations suggest that Ge-Ge bonds can be found already in $GeS_3$-15Ag (figure 5). The Ge-Ge coordination number is around 0.7±0.3, the concentration dependence is not significant.

It is to be noted that the S-S coordination number is around 1 and on the average each sulfur atom takes part in ~2 covalent bonds. The existence of S-S bonds was also tested by dedicated runs. It was found that raising the S-S cut off from 1.9 Å to 2.8 Å only slightly deteriorates the neutron diffraction and Ge K-edge EXAFS fits but drastically influences the fit of Ag K-edge EXAFS data as demomstrated in figure 6. (The reason for this change is that the elimination of S-S bonds brings about a forced increase of the Ag-S coordination number.) The total coordination number of sulphur increases from about 2 in $GeS_3$ glass to 3.30±0.4 in $GeS_3$-25Ag glass.

*4.3. Comparison of $AsS_2$-25Ag and $GeS_3$-25Ag glasses*

Since Ge-S-Ag and As-S-Ag glasses have similar potential applications it may be interesting to compare the environment of silver atoms and the changes of the host covalent matrices induced by alloying. The structure of glassy $AsS_2$-25Ag was investigated by diffraction techniques and EXAFS in a similar way [17]. It was found that the mean Ag-S coordination number is 3.34±0.4, while the Ag-Ag coordination number is 0.78±0.4. On the average, the total coordination number of Ag is close to 4. The Ag-S distance is practically the same in the two glasses, but the first peak is more pronounced in $AsS_2$-25Ag (figure 7).The Ag-Ag distance is 2.92±0.03 Å, which is close to the value found in $GeS_3$-25Ag (~2.96-3.01 Å). The coordination number of As is close to 3 in $AsS_2$-25Ag, and no As-As bonds had to be allowed to get reasonable fits. The S-As coordination number is around 1.5, while the S-S coordination number is not higher than the sensitivity of our approach (~0.3).

Thus, while Ge-Ge bonds are formed in $GeS_3$-25Ag, $AsS_{3/2}$ units remain intact in $AsS_2$-25Ag. The other important difference between the two glasses is that due to the nonvanishing S-S bonding each sulfur atom participates in ~2 covalent bonds in $GeS_3$-25Ag. The same number is around 1.5 in $AsS_2$-25Ag. It can be concluded that even if the chemical short range order is changed (due to the formation of Ge-Ge bonds) the connectivity of the covalent network of Ge and S atoms is not altered by the addition of Ag.

**5. Conclusions**

Short range order in $GeS_3$ and $GeS_3$-Ag glasses (15, 20 and 25 at.% Ag) was investigated by neutron and X-ray diffraction as well EXAFS measurements at the Ge and Ag K-edges. Structural models were obtained by fitting experimental datasets simultaneously with the reverse Monte Carlo simulation technique. It was found that S-S bonding is preserved even in $GeS_3$-25Ag and each S atom takes part in ~2 covalent bonds. The average coordination number of sulphur is 2.64±0.3 in $GeS_3$-15Ag while it is 3.30±0.3 in $GeS_3$-25Ag glass. Ge-Ge bonds appear already in $GeS_3$-15Ag composition; thus addition of silver changes chemical short range order of the host $GeS_3$ matrix. On the other hand, as each sulfur atom has ~2 Ge/S neighbours, the connectivity of the covalent network of Ge and S atoms is not altered by the addition of Ag. The average coordination number of Ag is 2.90±0.3 in $GeS_3$-15Ag glass and 3.79±0.3 in $GeS_3$-25Ag glass.

**Acknowledgments**

P.J. was supported by OTKA (Hungarian Basic Research Fund) Grant No. 083529. The authors thanks to project CZ.1.07/2.3.00/20.00254 "Research Team for Advanced Non-crystalline Materials" realized by European Social Fund and Ministry of Education, Youth and Sports of The Czech Republic within The Education for Competitiveness Operational Programme for financial support.

Table 1. Mass density and number density of the GeS$_3$-Ag glasses investigated. Accuracy of the mass density is about 0.15%.

| Composition | GeS$_3$ | GeS$_3$-15Ag | GeS$_3$-20Ag | GeS$_3$-25Ag |
|---|---|---|---|---|
| Density (g/cm$^3$) | 2.588 | 3.628 | 4.031 | 4.347 |
| Number density (atom/Å$^3$) | 0.03692 | 0.041968 | 0.043864 | 0.044654 |

Table 2. Minimum interatomic distances (in Å) used in the reverse Monte Carlo simulation runs.

| Composition | Atomic pair | | | | | |
|---|---|---|---|---|---|---|
| | Ge-Ge | Ge-S | Ge-Ag | S-S | S-Ag | Ag-Ag |
| GeS$_3$ | 3.1 | 2.0 | – | 1.95 | – | – |
| GeS$_3$-Ag | 2.35 | 2.0 | 3.15 | 1.95 | 2.2 | 2.85 |

Table 3. Coordination numbers of GeS$_3$-Ag glasses obtained by the simultaneous fitting of diffraction and EXAFS datasets.

| Composition | Coordination number | | | | | | | | |
|---|---|---|---|---|---|---|---|---|---|
| | $N_{GeGe}$ | $N_{GeS}$ | $N_{SGe}$ | $N_{SS}$ | $N_{SAg}$ | $N_{AgS}$ | $N_{AgAg}$ | $N_S$ | $N_{Ag}$ |
| GeS$_3$ | – | 3.90 | 1.30 | 0.86 | – | – | – | 2.16 | – |
| GeS$_3$-15Ag | 0.64 | 3.20 | 1.07 | 1.06 | 0.51 | 2.17 | 0.73 | 2.64 | 2.90 |
| GeS$_3$-20Ag | 0.69 | 3.30 | 1.10 | 0.96 | 0.95 | 2.86 | 0.82 | 3.01 | 3.68 |
| GeS$_3$-25Ag | 0.75 | 3.25 | 1.08 | 0.95 | 1.27 | 2.86 | 0.93 | 3.30 | 3.79 |

Table 4. Nearest neighbour distances (in Å) calculated from the models obtained by reverse Monte Carlo simulation

| Composition | Mean interatomic distance | | | | |
|---|---|---|---|---|---|
| | $r_{GeGe}$ | $r_{GeS}$ | $r_{SS}$ | $r_{SAg}$ | $r_{AgAg}$ |
| $GeS_3$ | – | 2.23±0.02 | 2.06±0.02 | – | – |
| $GeS_3$-15Ag | 2.43±0.02 | 2.22±0.02 | 2.10±0.04 | 2.53±0.03 | 2.96±0.05 |
| $GeS_3$-20Ag | 2.46±0.02 | 2.22±0.02 | 2.10±0.04 | 2.57±0.03 | 3.01±0.05 |
| $GeS_3$-25Ag | 2.44±0.02 | 2.22±0.02 | 2.10±0.04 | 2.56±0.03 | 2.97±0.05 |

**Figure captions**

Figure 1. Comparison of experimental and RMC-model curves of GeS$_3$ glass.

Figure 2. Partial pair correlation functions of GeS$_3$ glass.

Figure 3. XRD, ND, Ge and Ag K-edge measurements and RMC-model curves of GeS$_3$-25% Ag glass obtained by fitting the four datasets simultaneously.

Figure 4. Partial pair correlation functions of GeS$_3$-25Ag glass.

Figure 5. Comparison of the fits of Ge K-edge EXAFS data for the GeS$_3$-15Ag glass with and without Ge-Ge bonds (*symbols*: experimental data, *solid line*: fit).

Figure 6. Comparison of the fits of Ag K-edge EXAFS data for the GeS$_3$-25Ag glass with and without S-S bonds (*symbols*: experimental data, *solid line*: fit).

Figure 7. Comparison of the Ag-S partial pair correlation functions in GeS$_3$-25Ag (*dashes*) and AsS$_2$-25Ag (*solid line*) glasses.

Figure 1.

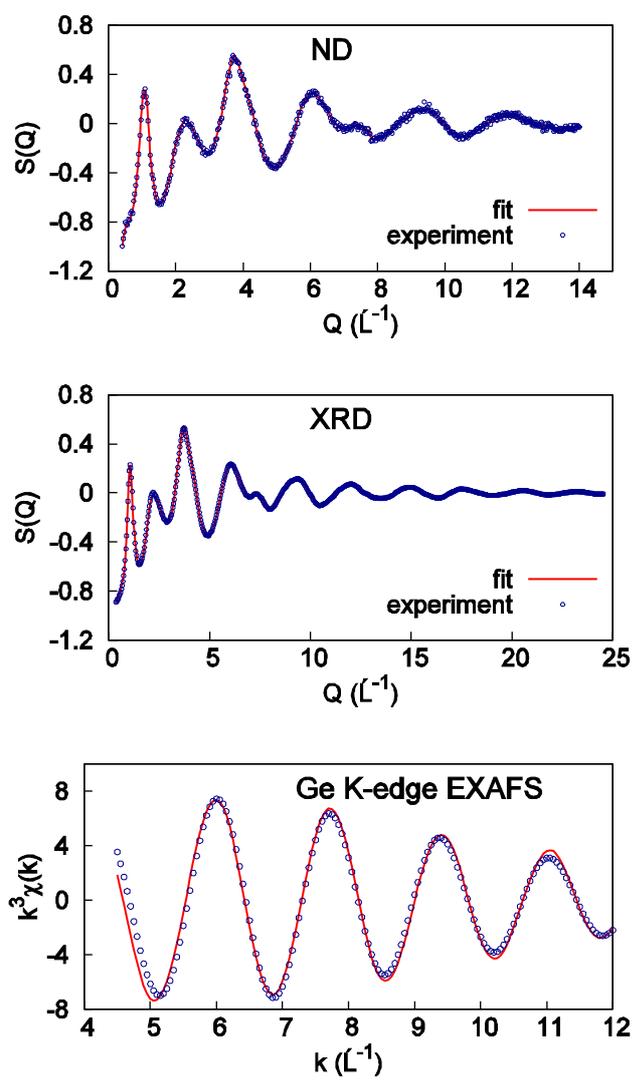

Figure 2.

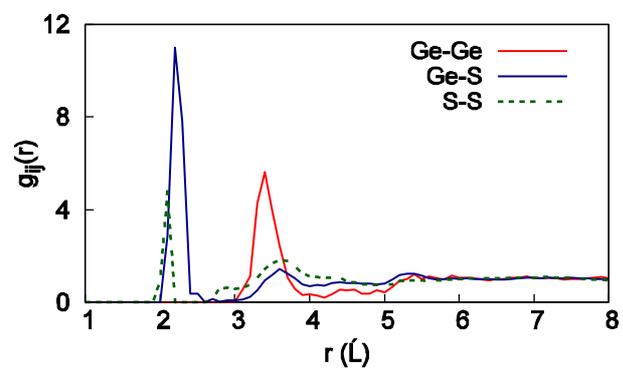

Figure 3.

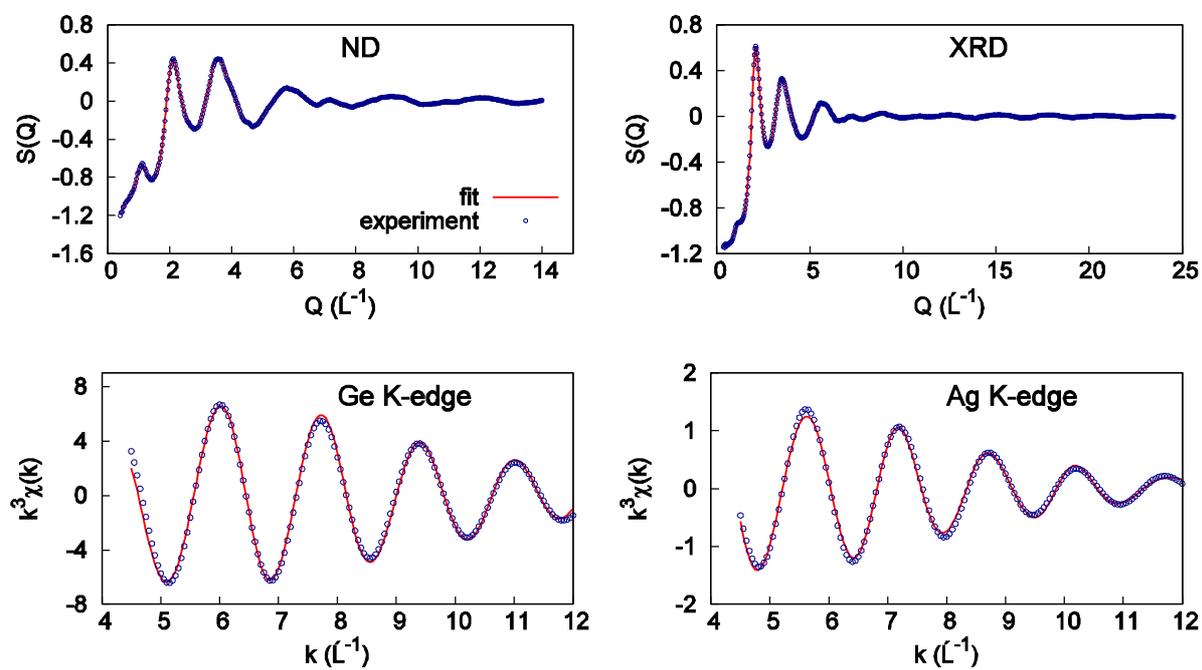

Figure 4.

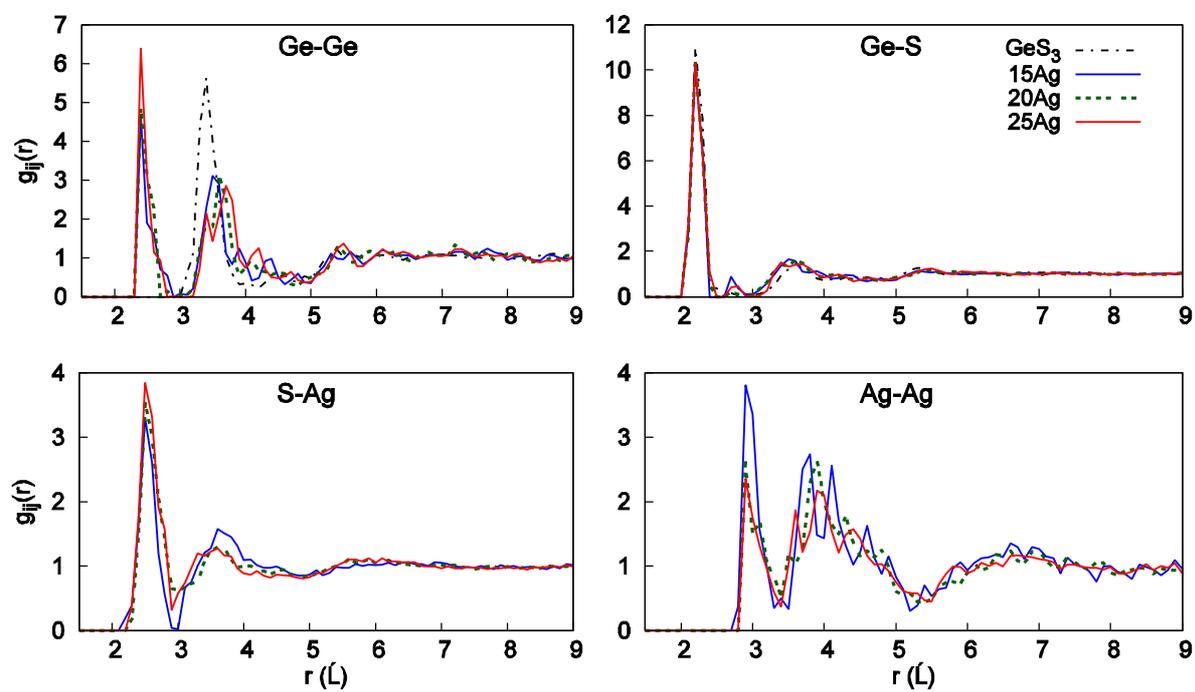

Figure 5.

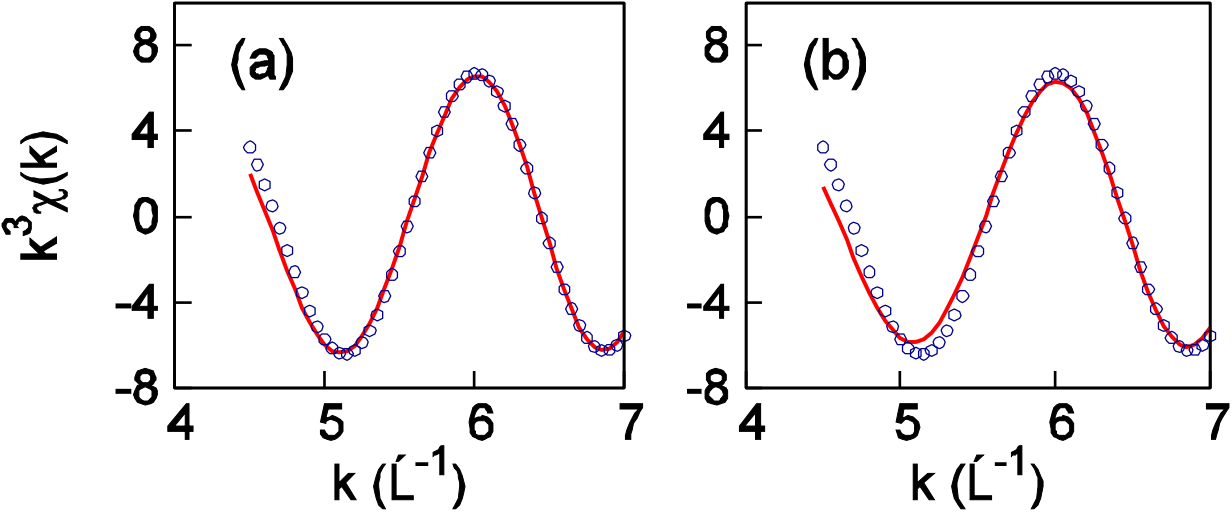

Figure 6.

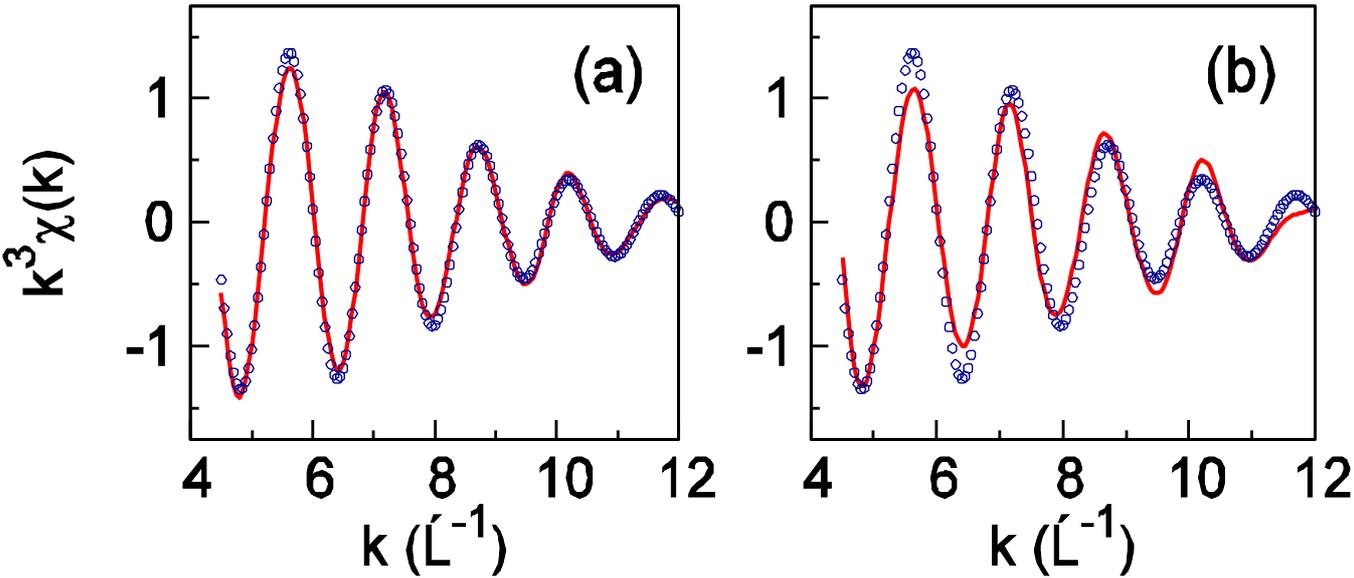

Figure 7.

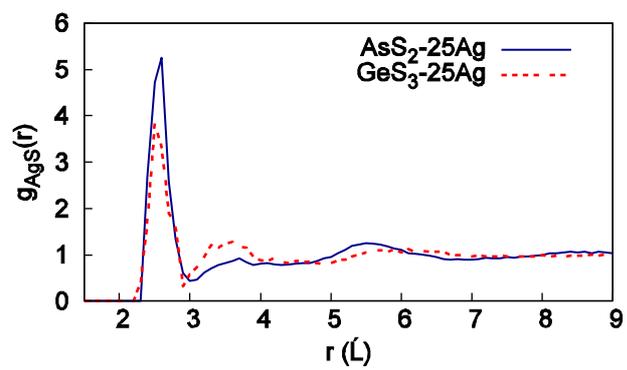